# Direct observation of electrically triggered Insulator-Metal Transition in $V_3O_5$ far below the phase transition temperature

Coline Adda[1*], Min-Han Lee[1,2], Yoav Kalcheim[1¥], Pavel Salev[1], Rodolfo Rocco[3], Nicolas M. Vargas[1], Nareg Ghazikhanian[1,2], Marcelo Rozenberg[3], and Ivan K. Schuller[1]

[1] Department of Physics and Center for Advanced Nanoscience, University of California-San Diego, La Jolla, California 92093, USA

[2] Materials Science and Engineering Program, University of California-San Diego, La Jolla, California 92093, USA

[3] Laboratoire de Physique des Solides, CNRS, Université Paris Saclay, 91405 Orsay Cedex, France.

[¥] present address Faculty of Materials Science and Engineering, Technion – Israel Institute of Technology, Haifa 32000, Israel

*Corresponding author: cadda@ucsd.edu

**Resistive switching is one of the key phenomena for applications such as nonvolatile memories or neuromorphic computing. $V_3O_5$, a compound of the vanadium oxide Magnéli series, is one of the rare materials to exhibit an insulator-metal transition (IMT) above room temperature ($T_c$ ~ 415 K). Here we demonstrate both static dc resistive switching (RS) and fast oscillatory spiking regimes in $V_3O_5$ devices at room temperature (120 K below the phase transition temperature) by applying an electric field. We use *operando* optical imaging to track a reflectivity change during the RS and find that a percolating high temperature metallic phase filament is formed. This demonstrates that the electrically induced RS triggers the phase transition. Furthermore, we optically capture the spiking oscillations that we link to the negative differential resistance regime and find the filament forms and dissolves via a periodic spatio-temporal instability that we describe by numerical simulations.**

Resistive switching in materials that undergo an insulator to metal phase transition (IMT) is the subject of intense fundamental research[1–5] and it could enable a variety of novel applications, ranging from optoelectronic devices,[6,7] oxide electronics,[8,9] to artificial neurons and synapses for hardware-level neuromorphic computing.[10–15] From the practical point of view room temperature operation of devices is highly desirable. The number of materials with the IMT above room temperature, however, is rather limited.[16] At present, most research focuses on resistive switching in $VO_2$ ($T_c \approx$ 340 K) and $NbO_2$ ($T_c \approx$ 1080 K). When resistive switching is realized at temperatures far below the phase transition, it is not always clear whether or how the switching is related to it. Several explanations for the origin of the resistive switching have been proposed: triggering of the IMT via Joule heating, doping the insulating phase with high electric field, and the so called "thermal runaway". In the first case, the electrical triggering of the IMT via Joule self-heating, leads to the formation of a high temperature phase metallic filament.[5,17] In the second case, the high electric field promotes charges from a low lying state into the conduction band thus electrically doping and destabilizing the Mott material.[3] "Thermal runaway" occurs by the formation of a low resistance filament at an intermediate temperature (that is below the high temperature metallic phase), which is possible because of a rapid decrease of resistance with increasing temperature in the insulating phase. It is rather well-established that the switching in pristine $VO_2$[3,5,17] at room temperature (i.e. ~50 K below $T_c$) can be driven by local Joule heating across the IMT. On the other hand, in $V_2O_3$ (and also radiation damaged $VO_2$), resistive switching can be accomplished by destabilization of the Mott state by charge injection.[3] In $NbO_2$, Joule heating has been claimed to the cause of both, local heating across the IMT[18,19] and "thermal runaway" [20–23] in the observation of resistive switching well below the Tc. This variety of possibilities illustrates that observation of resistive switching in a IMT material does not automatically indicate the triggering mechanism. It is therefore of critical importance to establish the physical driving force behind the switching, especially if the transition temperature is far from the room temperature.

Here we shall focus on the physical nature of the resistive switching in $V_3O_5$ thin films. Recently, resistive switching was demonstrated in $V_3O_5$ bulk crystals at room temperature (i.e. 120 K below

T$_c$).[24] It was not clear, however, whether this switching was produced by inducing the phase transition or by a thermal runaway in the insulating phase similar to NbO$_2$. By combining electrical transport measurements and direct optical imaging, we show that the resistive switching originates from a coupled thermo-electric triggering of the phase transition along a filamentary conductive path. This conclusion is further supported by numerical simulations that describe the physical mechanism in detail. Quite remarkably, we achieve control of the filamentary structure that we imaged in both, static I-V measurements and in a dynamic self-oscillating "spiking" regime. Furthermore, our results demonstrate that the phase transition could be induced even when operating at a temperature well below T$_c$, which is relevant for potential neuromorphic hardware applications.

The synthesis of pure phase V$_3$O$_5$ is challenging because the required stoichiometric range is very narrow ($VO_{1.666}$; $VO_{1.668 \pm 0.002}$).[25,26] In this work, we obtained V$_3$O$_5$ films by transforming V$_2$O$_3$ films in a high-temperature oxygen-rich environment. The details of the synthesis process are available elsewhere.[27] Two Ti/Au electrodes 20 µm wide and separated by a 20 µm gap were patterned on top of the film to make planar two terminal devices for resistive switching measurements. The planar device configuration enables an easy access to probe the optical response of V$_3$O$_5$ simultaneously with the electrical properties, which allowed us to detect the triggering of the phase transition during resistive switching.

We observed a clear signature of the phase transition in our V$_3$O$_5$, by using two independent techniques, resistance (R) and optical reflectivity (Ref) vs. temperature measurements (Figure 1a). Around T$_c$ ≈ 415 K, the resistance rapidly changes by about factor of three. At the same time, the devices display a significant reduction of the optical reflectivity. Both R(T) and Ref(T) show a continuous, non-hysteretic evolution near T$_c$, which is expected for a 2$^{nd}$ order phase transition.[28,29] There is an important difference, however, between the R(T) and Ref(T). R(T) shows a strong nonlinear (exponential) temperature decrease in the insulating state, below T < 400 K. Because of this, the relative resistance change from room temperature to the onset of the phase transition is comparable to the relative resistance change across the IMT. As a consequence, the observation of a resistance change during current induced resistive switching in a room temperature experiment, does not necessarily imply the electrical triggering of the phase transition has been achieved, which requires to overcome Tc. In contrast, since the optical reflectivity is nearly independent of temperature in the insulating state and shows a change only at the onset of the IMT, the observation of a change in the reflectivity would present a definitive evidence for the electrical triggering of the phase transition at the resistive switch.

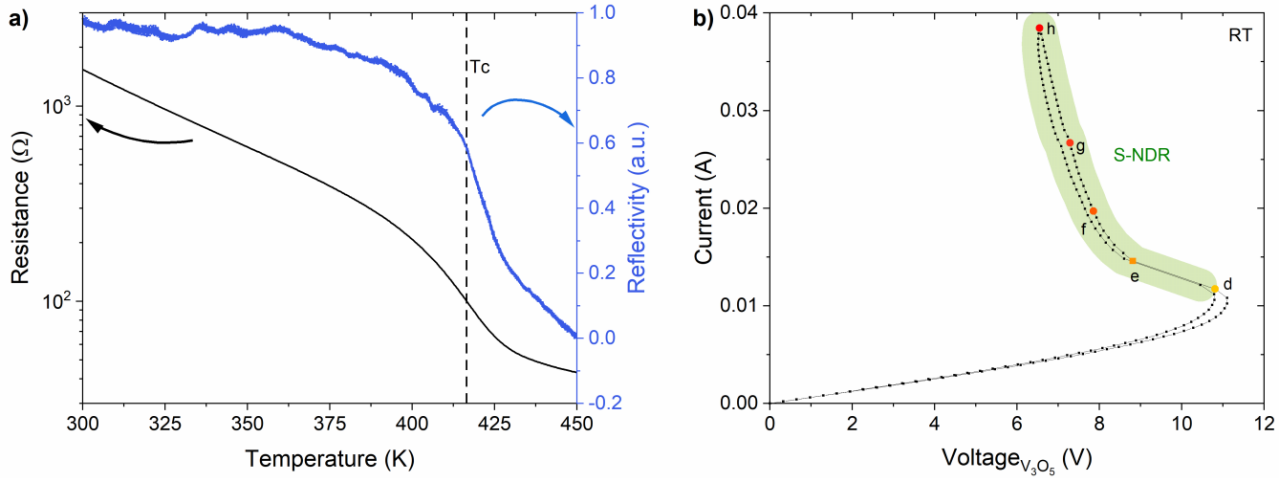

Figure 1: **Transport and reflectivity characteristics of a $V_3O_5$ device.** a) The resistance vs. temperature (R(T)) (in black) and the reflectivity vs. temperature (Ref(T)) (in blue) curves clearly show an insulator to metal transition (IMT) around 415 K. No thermal hysteresis can be observed in either curves indicating a 2$^{nd}$ order transition. b) Room temperature (RT) current - voltage characteristic. Above an ~ 12 mA threshold current, the $V_3O_5$ device switches into a low resistance state, through an S-shaped-NDR (dV/dI <0) regime.

$V_3O_5$ devices showed a pronounced resistive switching at room temperature. We observed (Figure 1b) that above a certain threshold, the current-controlled I-V characteristics develops an S-type negative differential resistance (S-NDR), [30] i.e. a region where the slope, dV/dI, is negative. The NDR corresponds to the onset of the resistive switching: as the current increases, the voltage drop across the device becomes smaller, indicating that the resistance suddenly decreases. The NDR first appears under a relatively large power, ~120 mW, dissipated in the device. Such large power is expected to cause a significant Joule self-heating, which most likely drives the observed resistive switching. However, as mentioned before, this observation alone does not permit to unambiguously determine whether the device has overcome the phase transition temperature. Another important observation is that the I-V characteristics does not show any significant hysteresis. This is in stark contrast with the previously reported measurements in $V_3O_5$ bulk crystals.[24] We attribute the absence of the I-V hysteresis in our devices to the thin film geometry.

The good thermal coupling between the Al$_2$O$_3$ substrate and the V$_3$O$_5$ film allows a fast steady-state temperature distribution in the device, that is established rapidly upon changing the applied voltage/current. The efficient thermal coupling in thin film devices allows fast speed operation, which we demonstrate next by probing the self-oscillation regime in V$_3$O$_5$.

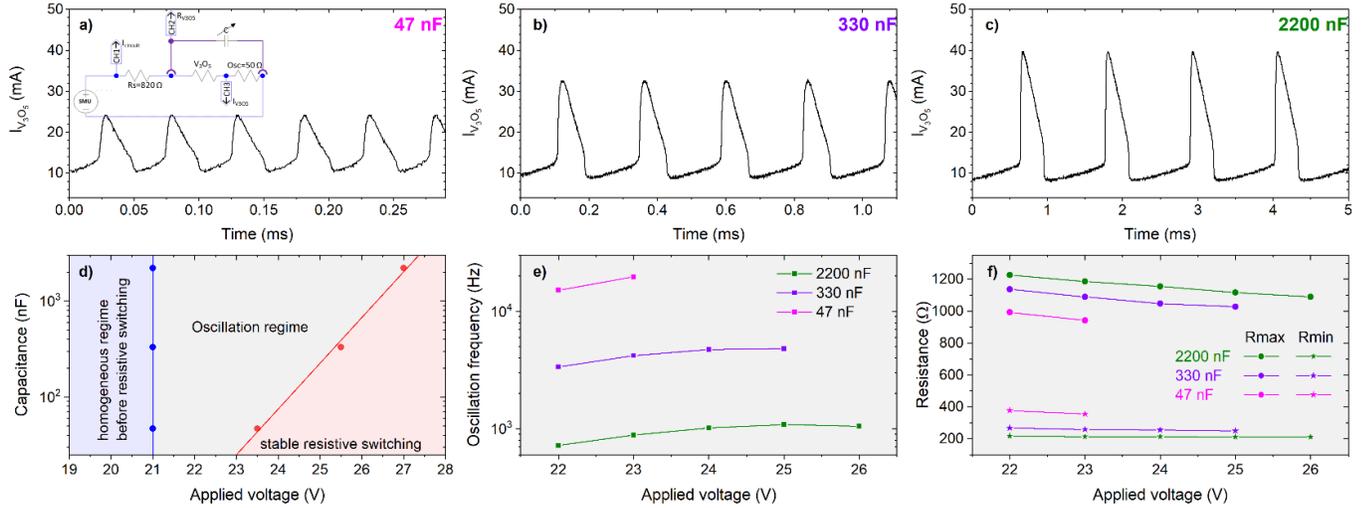

Figure 2: **Self-oscillation regime in V$_3$O$_5$.** a-c) Time dependence of the current through the V$_3$O$_5$ device for different capacitances: 47 nF in a), 330 nF in b), and 2200 nF in c). In inset, Pearson-Anson circuit. d) Summary of the three different regimes as function of the capacitance and the applied voltage. For voltages below 21 V (blue area), the resistive switching has not happened, and therefore the device is in a homogeneous regime, regardless of the capacitance. Above 21V, the device enters in an oscillation regime (grey area) that ends at higher voltages (red area) where the resistive switching state is stable. e) Oscillation frequency as a function of the applied voltage for different capacitance. The frequency decreases as the capacitance increases but increases with the voltage. Neither the homogeneous regime nor the stable resistive switching regime are shown. f) Minimum and maximum resistances of the V$_3$O$_5$ device during the oscillation regime as a function of the voltage and for different capacitances. Both the amplitude of the resistance change and the extrema resistance values depends on the capacitance. The maximum resistance decreases slowly with the applied voltage independently of the capacitance, the minimum resistance is nearly independent of the voltage.

Inducing dynamic self-oscillating resistive switching regime in IMT materials allows mimicking the spiking behavior of biological neurons, which could find applications in hardware-level implementation of neuromorphic computing.[31,32] We realized the self-oscillation regime in V$_3$O$_5$ by assembling a Pearson-Anson-type circuit (Figure 2a inset).[33] We connected a series resistor and a parallel capacitor to the V$_3$O$_5$ device and monitored the instantaneous voltage and current using an oscilloscope. We observed persistent, fast current oscillations at room temperature when the circuit is biased with a dc voltage (Figure 2a-c). The oscillation regime appears within a certain dc bias range that depends on the parallel capacitance value. The larger the capacitor, the wider the dc bias range where the oscillating regime can be induced (Figure 2d). The oscillation frequency also depends strongly on the capacitance, ranging from hundreds of Hz to several tens of kHz (Figure 2e). We also find a strong capacitance dependence for the amplitude and the extreme values of the self-oscillations (Figure 2f). As an example, the resistance change obtained with the 2200 nF capacitor is almost twice the one obtained with the 47 nF capacitor. Although the low resistance state is nearly independent of the applied voltage, the resistance in the high resistive state slightly decreases when increasing the applied voltage. As the voltage increases, the RC constant decreases in the insulating state (due to larger current densities and increased self-heating effect) leading to shorter oscillation periods. On the other hand, the relative fraction of time spent in the filamentary state (see SI) increases with voltage. Consequently, above a certain voltage the time spent in the filamentary state becomes equal to the oscillation period itself, and the oscillations stop, driving the system into a stable resistive switching regime. When comparing the extreme values of the self-oscillations to the dc I-V curve in Figure 1b, we find that the state of $V_3O_5$ alternates between points before the onset of NDR and points deep into the NDR region. This suggests that the two resistive switching modes, the static dc switching and the dynamic self-oscillating switching, originate from the same physical mechanism, which we investigate using coupled electrical and optical measurements.

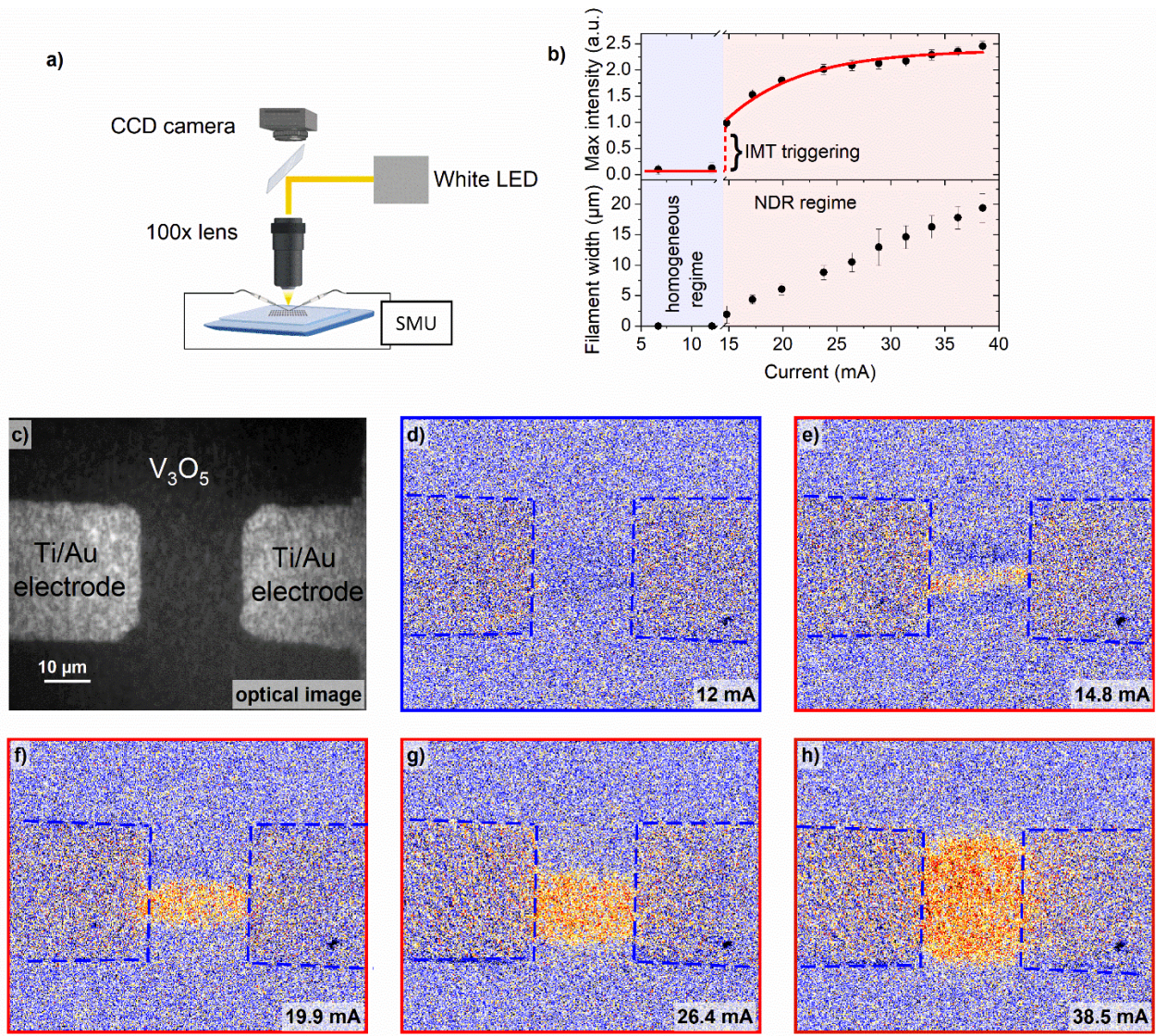

*Figure 3: **Optical imaging of the resistive switching in V$_3$O$_5$.** a) Measurement setup used to optically image the V$_3$O$_5$ device during resistive switching (created with BioRender.com). A white LED is used to illuminate the sample through a 100x objective. A CCD camera images the device as shown in c). b) Current dependence of the maximum intensity and the filament width. The light blue and red areas correspond to the regimes before and during NDR respectively. c) Optical image of the V$_3$O$_5$ device. The dark area is the V$_3$O$_5$ film, the two bright regions are the electrodes. The gap separating the two electrodes is ~20 µm. d- h) Differential contrast images of the resistive switching in V$_3$O$_5$ (see main text for details). The images show the formation of the filament that percolates between the electrodes. This filament is the high-temperature low-resistance phase of V$_3$O$_5$. d-h) correspond to 12 mA (yellow "d" point in Figure 1b)) , 14.8 mA (orange "e" point in Figure 1b)), 19.9 mA (dark orange "f" point in Figure 1b)), 26.4 mA (light red "g" point in Figure 1b)), and 38.5 mA (red "h" point in Figure 1b)) respectively.*

We observed clear evidence that voltage induced resistive switching in $V_3O_5$ is caused by the IMT and not by a "thermal runaway" higher conductance filament, which would hot but still within the insulating phase. This is found by simultaneously combining two independent techniques, optical reflectivity, and resistance measurement, during the resistive switching. We used an optical microscope to track the reflectivity changes while performing the switching measurements in current control mode (Figure 3a). To improve the signal to noise and contrast, we show the difference between the reference image recorded at zero current and the images acquired as the device is subject to various currents (Figure 3 d-h). This way we obtain differential images of the optical reflectivity change from the unperturbed state, which allows a reliable detection of the phase transition with spatial resolution.

No reflectivity change was found for the applied electric currents in 0 - 12 mA range, i.e. below the onset of the NDR (Figure 3d, light blue area in Figure 3b). However, as soon as the current drives the device into the NDR region, the reflectivity change shows an abrupt jump (Figure 3b) resulting in the appearance of a filament that percolates between the device electrodes (Figures 3 e-h). Filamentary switching is a very common phenomenon that has been observed in a variety of systems, [2,3,34–38] extending far beyond IMT materials. In our case, the important point is that the filament appears as the change of the optical reflectivity, which, according to our equilibrium R(T) and Ref(T) measurements (Figure 1a), implies that the IMT has been triggered during the resistive switching. Thus, we find that passing a sufficiently large dc current drives $V_3O_5$ into the high-temperature metallic phase even when the base experiment temperature is room temperature, i.e. 120 K below $T_c$ (295 K vs. 415 K).

By performing optical imaging of the $V_3O_5$ devices in the self-oscillation regime, we found evidence that the phase transition is triggered in each oscillation period. By using an exposure time of 55 ms, we were able to acquire images that correspond to temporal averages of 38-55 oscillation event per image (oscillation frequency of 0.7-1 kHz in the case of 2200 nF capacitor). When the applied bias drives the $V_3O_5$ device into the oscillation regime, a wide but a low-intensity filament appears in the images (Figure 4 b and c). Because this filament corresponds to the optical reflectivity change, we can immediately conclude that the $V_3O_5$ in the oscillation regime, undergoes the phase transition following the same argument as in the case of the static resistive switching (see the discussion in the previous paragraph). As the bias is increased, causing faster oscillation, the reflectivity change becomes more intense, but the filament width remains unchanged. When the bias becomes large enough to drive the device out of the oscillation regime and induces a stable resistive switching, the wide low-intensity filament abruptly collapses into a narrow high-intensity filament which continues to widen with further bias increase (Figure 4 d-f).

To identify the origin of the wide filament in the oscillation regime, we compared the filament intensity to the time fraction that the device spends in the low-resistance filamentary state (extracted from the time-dependency of the resistance measurements). We find a linear correlation between the intensity and the time (Figure 4g). This implies that the observed apparent wide low-intensity filament results from the temporal averaging done by our acquisition system. In fact, it results from the average of the rapid evolution of the system between two extreme states: the filamentary one when $V_3O_5$ has low resistance, and the homogeneous non-filamentary state when $V_3O_5$ recovers its high resistance. This conclusion is further supported by the fact that the width of the low-intensity filament during the oscillations is approximately the same as the filament width in the stable switching regime under the maximum current of the oscillations (compare Figure 4 c and e). Consequently, we find that despite the device base temperature being at 120 K below $T_c$, a filamentary phase transition in $V_3O_5$ is triggered in each oscillation cycle and the current discharge is intense and fast enough so to make the filament grow as wide as it would be in a stable dc switching under similar conditions.

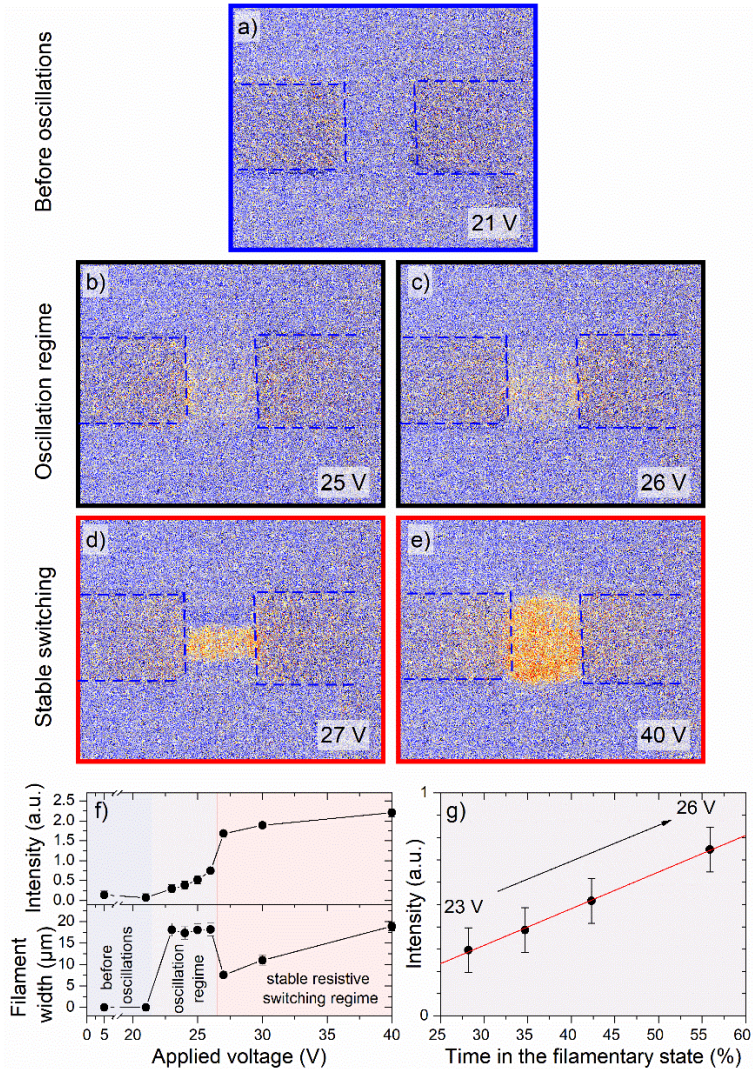

*Figure 4: **Optical imaging of the self-oscillation regime in $V_3O_5$.** a-e) Differential contrast images of the resistive switching in $V_3O_5$ acquired at the different voltages, corresponding to the homogeneous regime before oscillations in a), the oscillation regime in b) and c), and during the stable resistive switching regime after the oscillations in d) and e). f) Applied voltage dependence of the maximum intensity and width of the filament. The light blue, grey, and red areas correspond to the before oscillations, oscillations, and stable resistive switching regime, respectively. g) Intensity of the reflectivity change vs. time spent in the filamentary state during the oscillation regime. The time spent in the filamentary state is extracted from the electrical measurement for each voltage (see SI for details).*

The previous discussion is further supported by our model simulations. Based on the analysis of our numerical data, we find that the dynamics of successive creation and annihilation of the filament in the self-oscillation regime is governed primarily by the electrical "inertia" of the circuit, i.e., how fast the voltage/current builds up and discharges through in the $V_3O_5$ device. We simulated the same circuit as in the experimental part: a series resistor and a parallel capacitor connected to the $V_3O_5$ device. The elements and $V_3O_5$ device were simulated by means of a resistor network, which captures the filament formation. The physics of the system was modeled by using coupled time-dependent electrical and thermal equation described elsewhere.[39] Similar to the experiment, we found that under a certain voltage bias, the $V_3O_5$ device oscillates between low- and high-resistance states and that a wide filament, having a temperature above $T_c$, forms in every oscillation cycle (Figure 5). Upon closer examination of the simulated time dependences, two interesting features emerge. First, the maximum of the dissipated power (Figure 4d) lags behind the onset of filament formation (Figure 4, map γ). This lag is due to the capacitor discharge when the $V_3O_5$ switches in the low-resistance state. The discharge does not occur instantaneously due to the relatively large RC time constant of the circuit when the device is insulating. Eventually, the voltage on the capacitor reaches a threshold and its discharge produces a delayed power surge leading to the sudden formation of a very wide metallic filament in the $V_3O_5$, observed in the experiments (see Figure 3). Another interesting feature is the lag between the maximum dissipated power and the maximum temperature of the device (map ε). A similar lag can be also observed between the minimum power (map λ) and the minimum temperature (map α) (see SI). Those temperature lags are due to the thermal inertia of the device, i.e. how fast the device can heat up or cool down upon a sudden change of the dissipated power. Although the oscillation frequency in our model, as in the experiment, is mainly governed by the capacitor charging/discharging rate, there could be a situation when the RC time constant is small and the thermal inertia becomes the dominant factor determining the oscillation behavior. As long as the RC time constant is larger than the thermal time constant, the oscillation regime persists.[23] In our simulations, the temperature lag is about 20 times shorter than the oscillation period, therefore the oscillation frequency could be increased substantially by controlling the interplay between the electrical and thermal inertias.

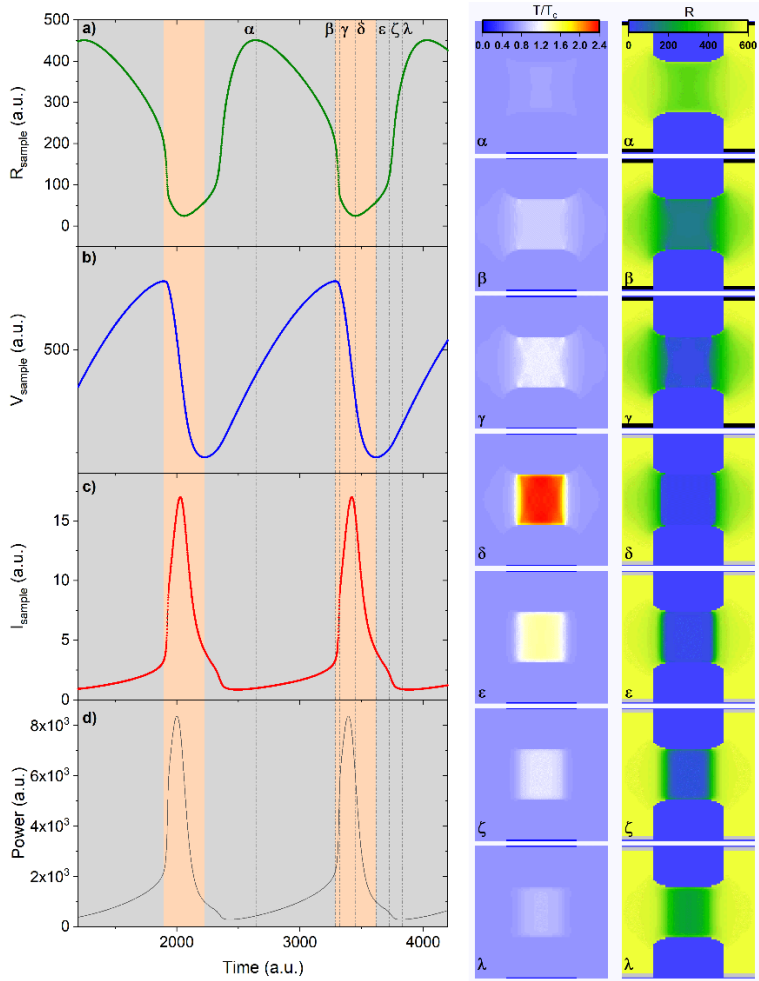

*Figure 5: **Simulation of the filament formation and dissolution during the oscillations.** Time dependence of the a) resistance, b) voltage, c) current and d) power simulated with a resistor network during the oscillation regime (see SI for details). The grey/orange backgrounds correspond to the charge/discharge of the capacitance. Resistance and temperature maps "α" correspond to the device in the insulating state. Maps "β" correspond to the initiation of the filament when the device starts to switch into a low resistance state. The filament starts to form, and the filament temperature reaches $T_c$. Maps "γ" correspond to the device in switched state when a high conductive filament is formed. The filament temperature is just above $T_c$. Maps "δ" correspond to the device in switched state at its minimum resistance, and when the maximum current is reached, just after the maximum power is reached. The filament has widen compared to maps "γ", and the filament temperature has increased. Maps "ε" correspond to the device in switched state when the capacitor is completely discharged. The resistance starts to increase as the filament narrows. The temperature is above $T_c$. Maps "ζ" correspond to the device in switched state but in a cooler state. The temperature, although still above $T_c$, is decreasing, and the filament continue to narrow. Maps "λ" corresponds to the device just before the relaxed state. The device has cooled down to $T_c$, and the filament has narrowed.*

In summary, we demonstrated that $V_3O_5$ exhibits volatile resistive switching both in static and in fast spiking oscillatory modes. This switching can be induced at room temperature, more than 120 K below the equilibrium IMT temperature of $V_3O_5$. Using a combination of transport and optical measurements, we showed that the resistive switching is triggered by the phase transition, which occurs via the formation of a percolating metallic-phase filament. The phase transition is triggered on a short time scale enabling the dynamics of filament creation and annihilation during each cycle in the fast oscillation regime. Because of the relatively low resistivity of $V_3O_5$ in the insulating state (~$10^{-2}$ Ω·cm) the phase transition can be driven far below $T_c$ without causing the device degradation. As a consequence, large power can be supplied to the device using moderate voltages which do not induce an electric field that is strong enough to initiate ion migration and/or dielectric breakdown. This feature may be important for applications, since $V_3O_5$ devices may have a better performance compared to $VO_2$, whose extremely large insulating state resistivity (>$10^1$ Ω·cm) often leads to the device breakdown before the electric triggering of the IMT. In addition, the $T_c$ of $V_3O_5$ is much lower compared to $NbO_2$, 420 K vs. 1080 K. This implies that the resistive switching in $V_3O_5$ has a substantially better energy efficiency than $NbO_2$. Overall, $V_3O_5$ displays a similar set of functionalities as $VO_2$ and $NbO_2$ (transitions in resistivity and optical reflectivity, resistive switching, fast spiking behavior), therefore using $V_3O_5$ could enable further improvement of a wide range of potential applications based on the IMT materials.

**Acknowledgement**

This collaborative work was supported as part of the "Quantum Materials for Energy Efficient Neuromorphic Computing" (Q-MEEN-C), an Energy Frontier Research Center funded by the U.S. Department of Energy, Office of Science, Basic Energy Sciences under Award # DE-SC0019273. RR is supported by French ANR "MoMA" project. The fabrication of the devices was performed at the San Diego Nanotechnology Infrastructure (SDNI) of UCSD, a member of the National Nanotechnology Coordinated Infrastructure, which is supported by the National Science Foundation (Grant ECCS-1542148).

# Direct observation of the electrically triggered Insulator-Metal transition in $V_3O_5$ far below the transition temperature – Supplementary Material

Coline Adda, Min-Han Lee, Yoav Kalcheim, Pavel Salev, Rodolfo Rocco, Nicolas Vargas, Nareg Ghazikhanian, Marcelo Rozenberg, and Ivan K. Schuller

**Part 1: Statistics on the oscillation regime**

As explained in the main text, the oscillation regime has been investigated for the oscillation regime and three different capacitors: 47 nF, 330 nF and 2200 nF.

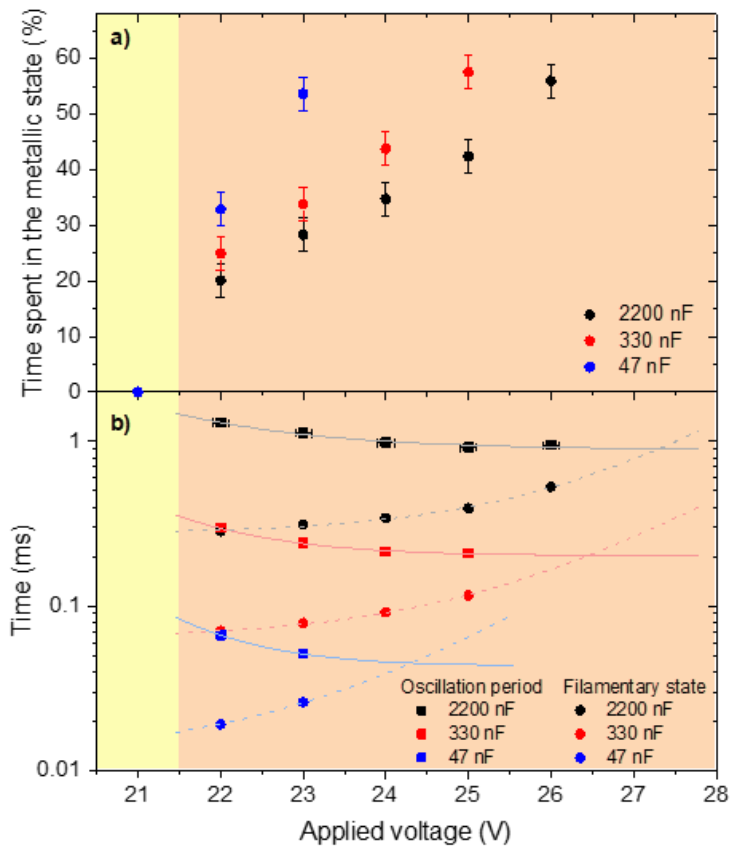

*Figure S 1: statistical study of the oscillation regime. a) Time spent in the filamentary state as function of applied voltage. The filamentary state was defined by a resistance below the threshold resistance of 750 Ω. b) Evolution of the period and time spent in the filamentary state with voltage. The lines going through the points are only a guide to help the visualization of an extrapolation. The stable resistive state regime starts at 23.5 V, 25.5 V and 27 V for the 47 nF, 330 nF and 2200 nF capacitors, respectively.*

In Figure S 1a, the time spent in the filamentary state has been extrapolated from the time dependence of the resistance. The filamentary state was defined by a resistance below the threshold resistance for the different capacitors. In Figure S 1b, the time spent in the filamentary state has been extrapolated from the time dependence of the resistance. The filamentary state was defined by a resistance below the threshold resistance. Since the resistances in the insulating state decreases with current, the period also goes down with current. On the other hand, the time spent in the filamentary state increases with current. Therefore, at some point, the filamentary time will catch up with the period. This marks the end of the oscillations and the beginning of the stable resistive switching regime. Figure S 2 shows that the minimum resistance achieved during the oscillations can be reached under higher voltage.

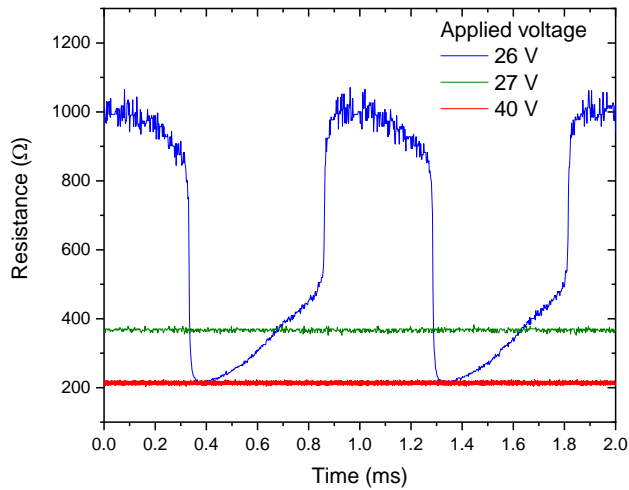

*Figure S 2: Time dependence of the $V_3O_5$ device resistance under different applied voltage. 26 V corresponds to the oscillation regime, 27 and 40 V correspond to the stable resistive switching regime. The minimum resistance achieved during the oscilations corresponds to the resistance of the device under 40 V.*

**Part 2: Analysis of the reflectivity images**

To help with the analysis of the reflectivity, the images have been processed as follow. After subtracting an image acquired at zero current from images acquired at different currents corresponding to various points on the I-V curve, we focused on the inter-electrode area as shown in Figure S 3a. The change of contrast intensity has been averaged along the Y-axis for each pixel along the X-axis. The results are displayed in Figure S 3b. The width of the electrodes is ~20µm. The beginning of the electrode has been set to 0.

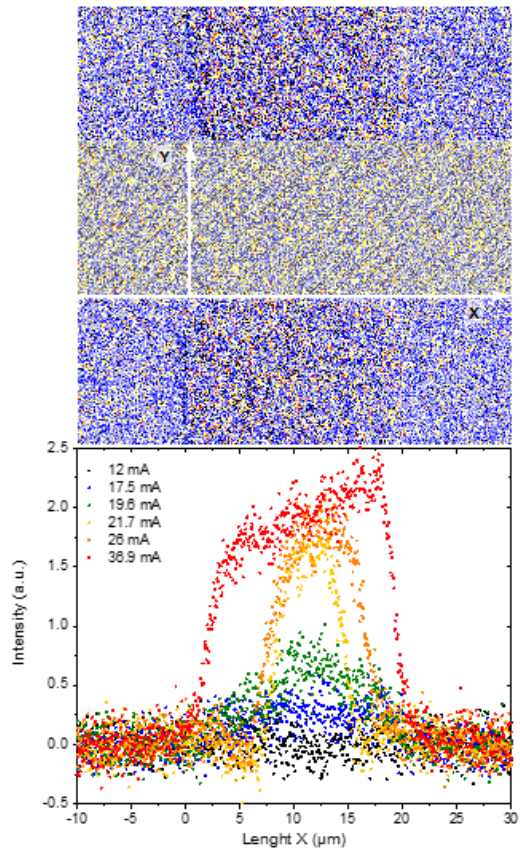

*Figure S 3: reflectivity analysis a) zoom into the gap area of images in main text. The yellow area is the area of interest. The contrast intensity has been averaged along Y-axis for each pixel along the X-axis  b) Average of the contrast intensity vs. Length X for different currents. The beginning of the electrode in a) has been set to 0.  The width of the electrode is 20 µm.*

**Part 3: Simulations**

To support our analysis, simulations based on the Mott resistor network model (first introduced in P. Stoliar *et al.* Adv. Mater. 25, 3222 (2013)), have been performed. In this model, the sample is described by a network of cells, each of which represents a nanoscale region of the sample and contains four resistors. The resistors connect the cells with one another and are characterized by a resistance that depends on the temperature in accordance with Figure S 3 a). When a voltage is applied across the network, currents flow through the resistors, generating heat. The temperature of the cells is given by the sum of the Joule heating contribution and the heat exchanged with the neighboring cells and the substrate with which the sample is in contact. Once the temperature of the cell has been computed, it is used to update its resistivity, i.e. the resistance of the four resistors within it. The system undergoes an insulator to metal transition when cells with low resistivity percolate from the top to the bottom of the network.  Figure S 4 corresponds to a simulation when no capacitor is added. The resistor network exhibits a I-V with a single resistive switching, and the thermal maps shows a creation of a small filament whose local temperature is above Tc at the resistive switching. As the current increases, the filament both widens and heats up.

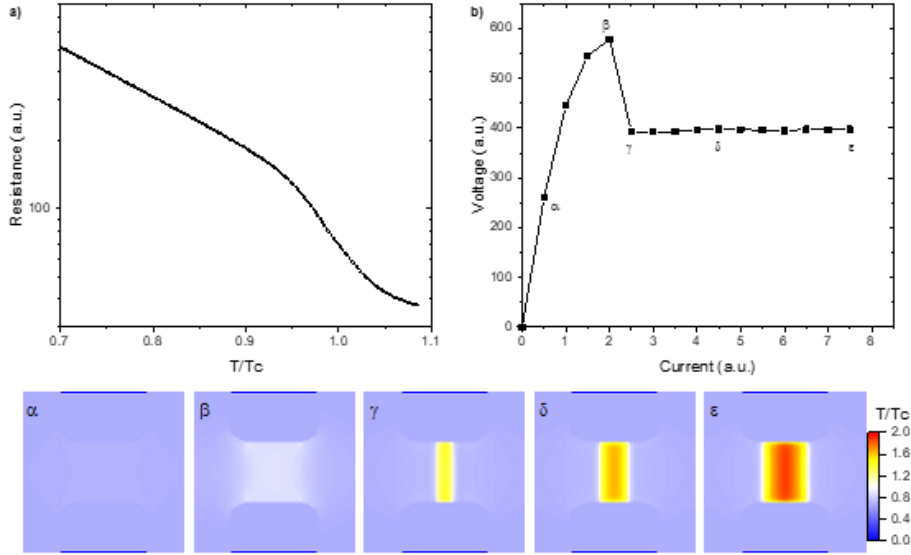

*Figure S 4: simulation of the filament formation. a) Resistance vs. temperature simulating a V3O5 R-T, with an IMT temperature $T_c$ = 415 K. b) simulated I-V. Thermal maps of the 2D resistor network are shown for different currents.*

Figure 5 of the main text corresponds to a resistor network with a capacitor added in parallel. An oscillatory regime is thus simulated. The resistance of the resistor network as well as the voltage and the current through it are shown. Grey parts correspond to the charge of the capacitor (Ic>0) and the light red parts corresponds to the discharge of the capacitor (Ic<0). Resistance and temperature maps have been calculated for different time during the oscillations. The filament forms by heating with a voltage threshold but then widens because of the discharge of the capacitor. Maps β are just before the electro-thermal filament formation. Maps ε are at the full discharge of the capacitor. From there, $V_{sample}$, which is also the voltage across the capacitor, start to increase. Therefore, input current goes to charge the capacitor. Between ε and λ, the current across the sample decreases but the voltage increases. In this phase, the filament cools down and start to disappear. This shows that heating/cooling rate of the sample is also important to consider. As shown in Figure S 5, the maximum temperature is reached after the maximal power.

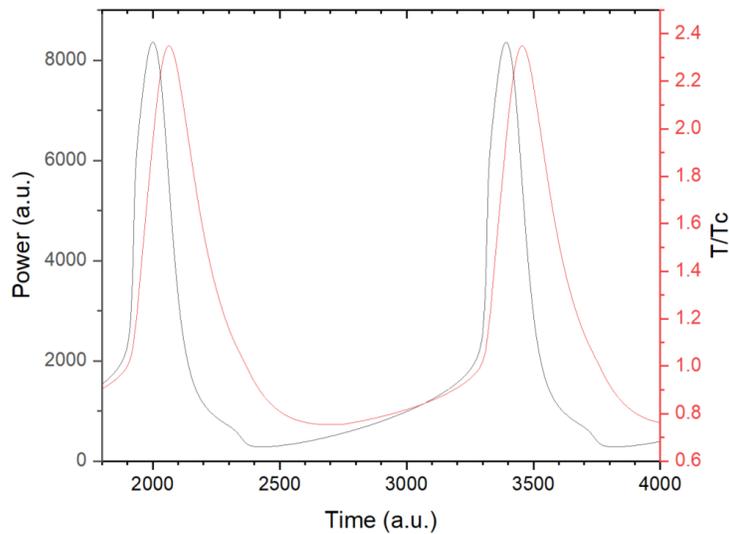

*Figure S 5: Time dependence of both the power and the temperature.*

**Part 4: Methods**

**Device fabrication**: In this work, $V_3O_5$ thin films are synthesized from 100 nm thick $V_2O_3$ thin film (grown by magnetron sputtering) on (100)$Al_2O_3$ substrate using a gas evolution technique.[28] By tuning very precisely the oxygen partial pressure and the temperature, it is indeed possible to control the oxygen stoichiometry of the thin film and therefore transforming the $V_2O_3$ into $V_3O_5$. 20 μm wide Ti/Au electrodes were patterned on top of the film separated by a 20 μm gap to create devices, so that an electric field triggered resistive switching can be induced.

**Electrical measurements:** Transport measurements were carried out using a Keithley 2450 Source Meter Unit. To develop spiking dynamics that are sought in neuromorphic applications, an oscillatory regime has been generated in the NDR region by adding a capacitor in parallel to the $V_3O_5$ device, as shown in the circuit in inset of Figure 2b. The oscillatory spiking behavior is measured as function of time, by using three channels of an oscilloscope. CH1 allows a measurement of the current through the series resistor ($I_{circuit}$) which give an oscillatory behavior for a certain voltage range (cf main text). These oscillations are used to determine the frequency (Figure 2e). CH2 provides the voltage across the capacitor, which allow to determine the resistance of the device. CH3 gives a measurement of the current going through the $V_3O_5$ device ($I_{V_3O_5}$), which give the spiking behavior (Figure 2a-c).

**Optical measurements:** The spatial dependence of the optical reflectivity of the $V_3O_5$ device was tracked as a function of applied voltage using an optical microscope. Figure 3a illustrates the setup used for the experiment. The device was illuminated with a white LED (400-700 nm) through a 100x lens of an optical microscope and imaged using a CCD camera.